\documentclass[twocolumn,preprintnumbers,showpacs,amssymb,amsmath,aps,prx]{revtex4-1}

\usepackage{tabularx}
\usepackage{float}
\usepackage{bm}
\usepackage{graphicx}
\usepackage{amsmath}
\usepackage{mathtools}
\usepackage{color}
\usepackage{mathrsfs}
\usepackage{hyperref}

\begin{document}

\title{Self-organizing maps as a method for detecting phase transitions and phase identification}

\author{Albert A.~Shirinyan,$^{1}$ Valerii K.~Kozin,$^{2,1}$ Johan Hellsvik,$^{3,4}$ Manuel Pereiro,$^5$ Olle Eriksson,$^{5,6}$ and Dmitry Yudin$^{7}$}
%\author{Valerii K.~Kozin$^{2,1}$}
%\author{Johan Hellsvik$^{3,4}$}
%\author{Manuel Pereiro$^5$}
%\author{Olle Eriksson$^{5,6}$}
%\author{Dmitry Yudin$^{7}$}
	
\affiliation {$^1$ITMO University, Saint Petersburg 197101, Russia \\ 
$^2$Science Institute, University of Iceland, Dunhagi-3, IS-107 Reykjavik, Iceland \\
$^3$Nordita, Roslagstullsbacken 23, SE-106 91 Stockholm, Sweden \\
$^4$Department of Physics, KTH Royal Institute of Technology, SE-106 91 Stockholm, Sweden \\
$^5$Department of Physics and Astronomy, Materials Theory Division, Uppsala University, Box 516, SE-75120 Uppsala, Sweden \\
$^6$School of Science and Technology, {\"O}rebro University, SE-701 82 {\"O}rebro, Sweden \\
$^7$Deep Quantum Labs, Skolkovo Institute of Science and Technology, Moscow 121205, Russia}
%\affiliation {$^2$Science Institute, University of Iceland, Dunhagi-3, IS-107 Reykjavik, Iceland}
%\affiliation {$^3$Nordita, Roslagstullsbacken 23, SE-106 91 Stockholm, Sweden}
%\affiliation{$^4$Department of Physics, KTH Royal Institute of Technology, SE-106 91 Stockholm, Sweden}
%\affiliation{$^5$Department of Physics and Astronomy, Materials Theory Division, Uppsala University, Box 516, SE-75120 Uppsala, Sweden}
%\affiliation{$^6$School of Science and Technology, {\"O}rebro University, SE-701 82 {\"O}rebro, Sweden}
%\affiliation{$^7$Deep Quantum Labs, Skolkovo Institute of Science and Technology, Moscow 121205, Russia}

\begin{abstract}
Originating from image recognition, methods of machine learning allow for effective feature extraction and dimensionality reduction in multidimensional datasets, thereby providing an extraordinary tool to deal with classical and quantum models in many-body physics. In this study, we employ a specific unsupervised machine learning technique---self-organizing maps---to create a low-dimensional representation of microscopic states, relevant for macroscopic phase identification and detecting phase transitions. We explore the properties of spin Hamiltonians of two archetype model systems: a two-dimensional Heisenberg ferromagnet and a three-dimensional crystal, Fe in the body-centered-cubic structure. The method of self-organizing maps, which is known to conserve connectivity of the initial dataset, is compared to the cumulant method theory and is shown to be as accurate while being computationally more efficient in determining a phase transition temperature. We argue that the method proposed here can be applied to explore a broad class of second-order phase-transition systems, not only magnetic systems but also, for example, order-disorder transitions in alloys.
\end{abstract}
\maketitle
	
{\it Introduction.} Recently, machine learning has been suggested as a tool to investigate many-body quantum systems \cite{Arsenault2014,Mehta2014,Torlai2016,Ohtsuki2016,Carrasquilla2017,Nieuwenburg2017,Carleo2017,Huang2017,Zhang2017,Saito2017,Ohtsuki2017,Tanaka2017,Hu2017,Schindler2017,Liu2017,Wetzel2017,Torlai2017,Broecker2017a,Chng2017, Broecker2017b,Costa2017,Nomura2017,Chng2018,Janusz2018,Torlai2018,Zhang2018,Fujita2018,Rao2018,Mills2018,Saito2018,Yoshioka2018}. In essence, machine learning deals with systems with an extremely huge number of degrees of freedom in data space rather than in phase space of quantum statistics. It is therefore not surprising that even networks of simple architecture can be trained by means of supervised learning to detect very peculiar phases in a variety of systems \cite{Carrasquilla2017,Nieuwenburg2017}, including topological and many-body localization phase transition \cite{Schindler2017}. The technique relies on sampling a physical system in a weighted way, and projecting the data onto hidden layers which filter out the irrelevant local fluctuations in the system, leaving only the large-scale behavior determining the macroscopic properties \cite{Mehta2014,Janusz2018}. Further, it was demonstrated \cite{Carleo2017} that restricted Boltzmann machines can be used to formulate a very efficient many-body wavefunction ansatz depending on a relatively small number of parameters even for a large number of spins ($\sim$10$^2$), self-adjusting via gradient descent-based reinforced learning. This allowed for computing both ground states and dynamically evolved states of large many-body systems, with excellent accuracy. Furthermore, this algorithm has been generalized to bosonic and fermionic Hubbard models \cite{Saito2017,Nomura2017}. The application of machine learning to quantum-information problems in condensed matter physics has also received significant interest recently, opening avenues for the direct experimental observation of the entanglement entropy \cite{Torlai2018}.
    
In the meantime, there is a growing interest toward a versatile methodology that, on one hand, reduces the dimensionality of the data space, while preserving its topology on the other. In this Rapid Communication we propose a method for determining phase transitions which, in contrast to the previous studies handling Ising-like models, makes it possible to associate the symmetry breaking during a second-order phase transition with a noticeable change in the topology of a certain space. We construct this target space based on a number of microscopic states generated with Monte Carlo simulations for two archetypal examples, namely, a two-dimensional ferromagnet on a square lattice (2DFM) and bcc iron (bcc Fe). We further apply an unsupervised machine learning method in the form of {\it self-organizing, or Kohonen, maps} (SOM) and compare the obtained results with those from cumulant method theory. We show that SOMs are able to correctly produce relevant two-dimensional representation of microscopic states, which allows one to visually observe symmetry breaking through a phase transition. The machine learning algorithms proposed here allow for a direct way for determining the critical temperature, while an intuitive interpretation of phase transitions in terms of principal component analysis (PCA) is also possible.

{\it Model systems.} For a vast class of magnetic compounds the microscopic description with high level of accuracy can be achieved within the Heisenberg exchange model. Whereas the Mermin-Wagner theorem establishes that an isotropic Heisenberg spin system in two dimensions cannot have a long-range ordering, the addition of anisotropy to the model changes the situation. The case of easy-plane exchange anisotropy is commonly referred to as the XXZ model, a system which akin to the XY model can display a Berezinskii-Kosterlitz-Thouless transition \cite{Cuccoli1985}. With easy-axis exchange, or single-ion, anisotropy, the system exhibits a second-order phase transition \cite{Leonel2006}. In the following we consider 2DFM on a square lattice with single-site easy-axis anisotropy, as well as the Heisenberg model for Fe in its ground state crystal structure (bcc). For 2DFM we consider a square lattice of classical spins with edge length $L$, corresponding to a total number $N=L\times L$ of sites, and use periodic boundary conditions. The nearest-neighbor Heisenberg exchange is assumed to have a typical strength for transition-metal systems, $J=1$ mRyd, and the parameter of easy-axis anisotropy is $K_\mathrm{anis}=0.2$ mRyd (see Supplemental Material, Sec.~A \cite{supp}). For bcc Fe, one may neglect the tiny magnetocrystalline anisotropy, and the spin Hamiltonian is of pure Heisenberg type. However, the exchange coupling parameters are long ranged, and have here been obtained from first-principles electronic structure calculations. Including up to the fourth coordination shell, we use the same set of exchange couplings $J_1=1.3377$ mRyd, $J_2=0.7570$ mRyd,  $J_3=-0.0598$ mRyd, and $J_4=-0.0882$ mRyd as calculated and used in Refs. \cite{Tao2005,Skubic2008} for a $L \times L \times L$ conventional bcc lattice with periodic boundary conditions as a simulation cell, corresponding to a number $N=2L^3$ of spins (see Supplemental Material, Sec.~A \cite{supp}).

In general, characterizing a phase transition requires a proper identification of the temperature point where the order parameter $M(T)$ goes to zero. For ferromagnetic Heisenberg-like models the order parameter is defined as the average magnetization per spin and in finite size systems the latter is not sharp enough at high-temperature regime where the role of fluctuations becomes important. To subdue this limitation and correct for finite size scaling one can apply the cumulant crossing method \cite{Binder1981a,Binder1981b}. Direct application of this approach to second-order phase transitions suggests that for Ising-like models in the thermodynamic limit the Binder cumulant $U(T)\rightarrow 0$ for $T>T_c$, whereas $U(T)\rightarrow2/3$ for $T<T_c$ as the lattice size increases \cite{Binder1981a}. For large enough systems $U(T)$ for different lattice sizes cross at a fixed point which can be identified with the critical temperature.
    
{\it Dataset.} To generate the appropriate spin configurations, we employed Monte Carlo simulations, with a heat bath algorithm for Heisenberg spin systems~\cite{Olive1986} as implemented in the \small{UPPASD} software~\cite{Skubic2008}. We use  $10^5$ Monte Carlo steps for equilibration, and $10^6$ Monte Carlo steps for the measurement phase. A sampling interval of ten steps was used for averages, susceptibility, total energy, and Binder cumulant measurements. Moreover, a sampling interval of 1000 (or 10 000) steps was used for snapshots of the whole spin configuration, resulting in up to 1000 (or 10 000) snapshots for each system and size. The results of these simulations are presented in the Supplemental Material, Sec.~A \cite{supp}. We note here, however, that they show a phase transition between ferromagnetic and paramagnetic phases for both 2DFM and bcc Fe. The critical temperature for the 2DFM is estimated from Binder cumulants to be around $T_c\approx 222$~K and for bcc Fe $T_c\approx 915$~K. For the machine learning method, we use as our training set data from the cells with edge length $L=80$, $L=120$ and $L=200$ (with the total number of spins being 6400, 14 400 and 40 000, respectively) for the 2DFM, and $L=24$, $L=28$, $L=36$ (with 27 648, 43 904, 93 312 spins) for bcc Fe.
 
\begin{figure}[h!]
    \includegraphics[width=\linewidth]{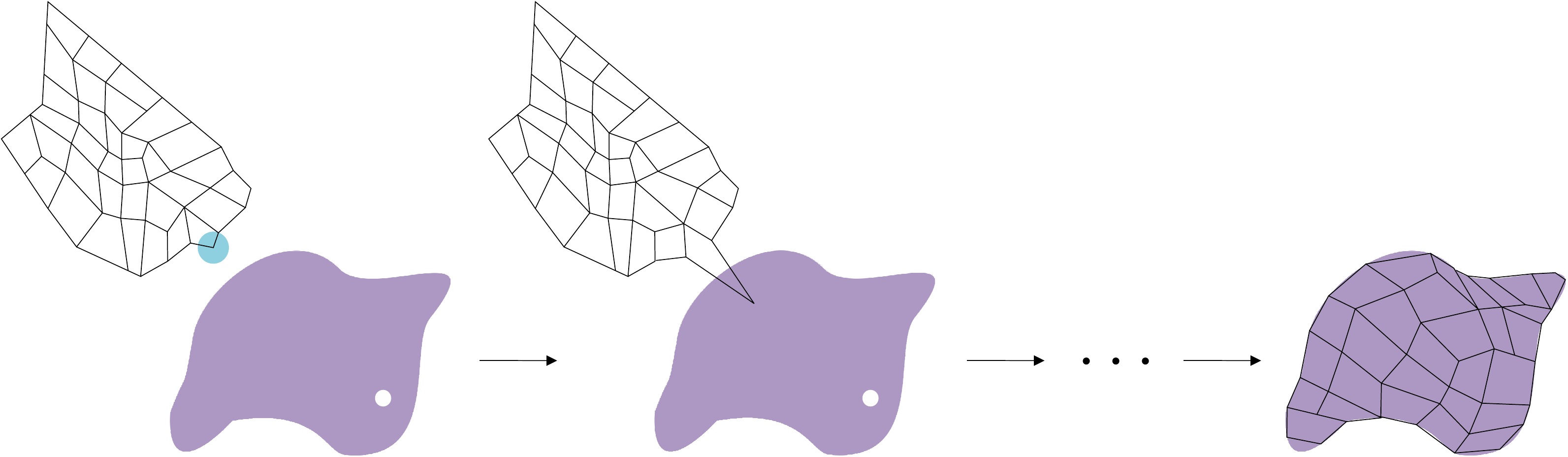}
    \caption{Training of a self-organizing map (SOM). The distribution of the training data is depicted by a blob, and the small white dot is the current training data chosen from that distribution. At first (leftmost) the SOM nodes are arbitrarily located in the data space. The highlighted node which is nearest to the training data is selected. It is moved toward the training data, as are its neighbors on the grid (but to a lesser degree). After many iterations the grid tends to approximate the data set (rightmost).}
    \label{fig:Kohonen_work}
\end{figure}

{\it Ideology of SOM.} A SOM, first introduced by Kohonen~\cite{Kohonen1982a,Kohonen1982b}, represents a neural network that performs visualization and clusterization by projecting a multidimensional space onto a lower dimensional one (most often, two-dimensional), and is trained using unsupervised learning. A SOM consists of components called nodes, or neurons, whose number is specified by the analyst. Each node is described by two vectors: the first one is the so-called weight vector, $\bm{w}$, of the same dimension as the input data, and the second vector, $\bm{r}$, is the one which gives the coordinates of the node on the map. The Kohonen map is visually displayed using an array of rectangular or hexagonal cells, associated with the respective node. During the training process, depicted schematically in Fig.~\ref{fig:Kohonen_work}, the weight vectors $\bm{w}(\bm{r})$ of the nodes approach the input data: for each observation (sample), the node with the closest weight vector is chosen, and its value moves toward the sample, together with the weight vectors of several neighbor nodes. The update formula for a weight vector $\bm{w}(\bm{r})$ is 
\begin{equation}
    \label{eq:SOM_training}
    \bm{w}_{n+1}(\bm{r})=\bm{w}_{n}(\bm{r})+\theta_n(\bm{r}',\bm{r})\cdot \alpha_n\cdot\left[\bm{d}_m-\bm{w}_{n}(\bm{r})\right],
\end{equation}
where $n$ is the step index, $m$ stands for an index in the training set, $\bm{d}_m$ is the sample vector, $\bm{r}'$ denotes the coordinates of the node with the closest weight vector to the $\bm{d}_m$, and $\alpha_n$ is a monotonically decreasing learning coefficient. In Eq.~(\ref{eq:SOM_training}), $\theta_n(\bm{r}',\bm{r})$ is the neighborhood function which depends on the grid distance between the neurons at $\bm{r}'$ and $\bm{r}$. In the simplest form it equals 1 for all neurons close enough to $\bm{r}'$ and 0 otherwise, though the Gaussian function could be an alternative option (regardless of the functional form, the neighborhood function shrinks as $n$ increases). Thus, if two observations are close in the set of input data, they would correspond to nearby nodes on the map. The repeating training process, enumerating the input data, ends when the SOM reaches an acceptable error (predetermined by the analyst), or if a specified number of iterations is done. As a result, the SOM classifies the data into clusters and visually displays the multidimensional input onto a two-dimensional plane, relating the vectors of similar characteristics to neighboring cells (an illustration of training process is shown in Fig.~\ref{fig:Kohonen_work}).
    
{\it Constructing a target space.} To apply the methods described above, one has to correctly organize the input data into a target space (in our case, the spin states of the system obtained from Monte Carlo simulations). Importantly, we demand an internal geometry that reflects the regularities that are of interest. The results of Monte Carlo simulations provide us with an array of $K$ spin states for each selected temperature at which they have been extracted (a higher number of states were generated in the vicinity of the phase transition temperature), and these states are represented by $3N$-dimensional vectors, where $N$ is the total number of lattice sites multiplied by three projections (in $x$, $y$, and $z$ directions) of the spins. For convenience, we reshape this array of states into a rectangular $K\times3N$ matrix, 
\begin{equation}
    \label{eq:targ_space}
    S = \begin{pmatrix}
    s^{1,1}_{x} & s^{1,1}_{y} & s^{1,1}_{z} & \dots  &s^{N,1}_{x} & s^{N,1}_{y} & s^{N,1}_{z}\\
    \vdots  &         &         & \ddots &        &     &        \\
    s^{1,K}_{x} & s^{1,K}_{y} & s^{1,K}_{z} & \dots  &s^{N,K}_{x} & s^{N,K}_{y} & s^{N,K}_{z} \\
    \end{pmatrix},
\end{equation}
where $s^{i,k}_{j}$ is the $i$th-site spin projection on the $j$th axis in  simulation $k$.
    
The first and most obvious way to form the target space is to take the rows of the matrix $S$ as its elements, thus  obtaining for each temperature a set of $K$ $3N$-dimensional vectors (in other words, to use directly a set of spin states). However, the angular distribution of vectors from any of these sets is practically isotropic, since each spin state obtained using Monte Carlo simulations has a random direction of the average spin, and the only information that we can get by observing the modification of the geometry of such space, is a change of its diameter with temperature. A more practical geometry can be obtained by forming the target space out of the columns of this matrix. In the following, we show that if the system is in the magnetically ordered phase, a clustering of the vectors of the constructed target space takes place, whereas for the disordered phase this does not happen.
    
{\it Clustering of the target space.} Let us fix the temperature $T$, and consider Monte Carlo step $k$ of the sampling phase. We then consider two columns, $\bm{y}(i,j)$ and $\bm{y}'(i',j')$, of the matrix $S$,
\begin{equation}
	\bm{y}(i,j) = \begin{pmatrix}
    s^{i,1}_{j}\\
    \vdots\\
    s^{i,K}_{j} \\
    \end{pmatrix},~ 
    \bm{y}'(i',j') = \begin{pmatrix}
    s^{i',1}_{j'}\\
    \vdots\\
    s^{i',K}_{j'} \\
    \end{pmatrix}.
\end{equation}
To write down $\bm{y}$ and $\bm{y}'$ without their arguments, we also fixed the first site number $i$, its projection $j$, and do the same with $i'$ and $j'$. In this way, we have two vectors $\bm{y},\bm{y}'\in\mathbb{R}^K$, which are close, if the Euclidean distance between them is small enough relative to some characteristic value, which, in our case, should be the diameter of the target space. The physical meaning of this proximity is that the given projections of the corresponding lattice site states, described by these vectors, are close in each simulation.
    
If the vector characterizing the microstate of the system is calculated by averaging over the lattice, and its length increases during some process, then vectors, describing lattice site states become more codirectional, and vice versa, an increase in the proportion of relatively codirectional vectors over the lattice sites leads to an increase in the modulus of the microstate parameter. This obvious reasoning, together with the target space constructed above, forms the basis of the here proposed phase determination method: in the case of a high magnetization, for each simulation $k$, the projections of the lattice sites spins on the same axis ($j=j'$) are close for the majority of sites $i$, while such a proximity of different projections ($j \ne j'$) would mean that in a significant part of the observations, the average spin tends to some specific directions, contradicting the isotropic distribution of data, obtained by the Monte Carlo simulation, e.g., represented by the criterion,
\begin{equation}
    \label{eq:phase_cond}
	|y_k(i, j)-y'_k(i',j')|_{j=j'} < |y_k(\tilde{i}, \tilde{j})-y'_k(\tilde{i}',\tilde{j}')|_{\tilde{j} \ne \tilde{j}'},
\end{equation}
for all $k$, $j$, $j'$, $\tilde{j}$ and $\tilde{j}'$. Thus, we can expect that in the ferromagnetic phase, the vectors in the target space will be grouped into clusters, corresponding to the projections onto $x$, $y$, and $z$ axes, while in the paramagnetic phase such clustering should be absent because of the much more isotropic distribution of the lattice site spins. Noteworthy, an intuitively clear PCA-based \cite{Jolliffe2002} graphical analysis of the magnetic phase transition unambiguously demonstrates three (for bcc Fe) or two (for 2DFM) well-separated clusters in the target space in the ferromagnetic phase, which are merging together in the disordered phase. This is illustrated in the Supplemental Material, Sec.~B \cite{supp}.
    
\begin{figure*}[t]
    \includegraphics[width=\linewidth]{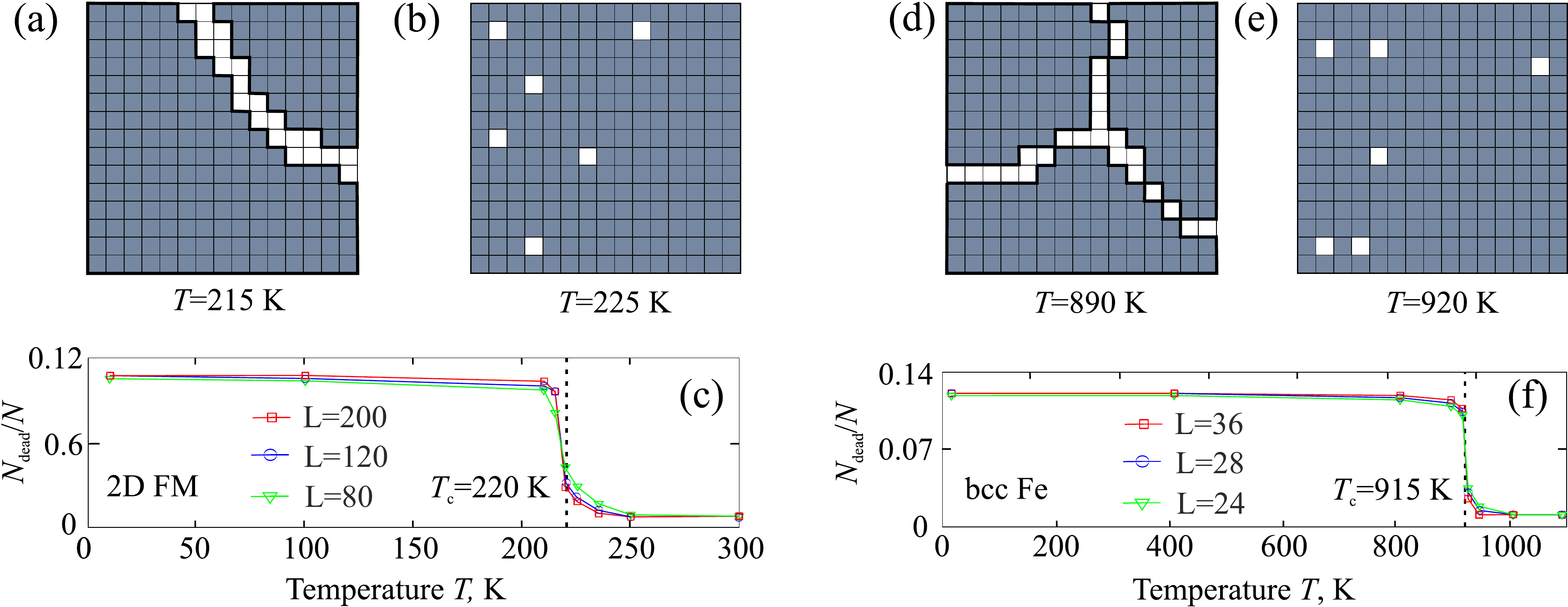}
    \caption{Neural activity in the SOM close to the phase transition. The gray cells represent the activated neurons, whereas the white ones mark the dead neurons. Below the critical temperature: at $T=215$~K for 2DFM, $L=120$ (a) and at $T=890$~K for bcc Fe, $L=28$ (d) the well-separated clusters of activated neurons are clearly visible. Above the critical temperature: at $T=225$~K for 2DFM, $L=120$ (b) and at $T=920$~K for bcc Fe, $L=28$ (e) no clusters are present. The ratio of the dead neurons to the total number of neurons, $N_\text{dead}/N$, for different lattice sizes as a function of temperature is shown for 2DFM (c) and bcc Fe (f). The sharp drop of $N_\text{dead}/N$ indicates the phase transition, revealing minor sensitivity to lattice size.}
    \label{fig:SOM}
\end{figure*}    

{\it Clustering detection by SOMs.} The advantage of using SOMs relies on the fact that they allow one to construct a two-dimensional projection of the multidimensional data distribution, while preserving the topology \cite{Graepel1997}. For this purpose, the target space vectors are normalized and centered, a SOM of a certain size is selected, a uniform distribution of the weight vectors of the nodes is set, and the SOM is trained according to  Eq.~(\ref{eq:SOM_training}). We observed best results for square maps of $\sim 15\times15$ nodes, providing thus the size for which cluster formation can be clearly distinguished (for more information about choosing the map size, see Supplemental Material, Sec.~C \cite{supp}). All nodes of the trained map can be divided into two types: those that are not activated even once during training process (so-called \textit{dead neurons}) and those that are activated at least once. In the case of the ferromagnetic phase, when Eq.~(\ref{eq:phase_cond}) is fulfilled, groups of activated nodes, separated by a band of dead neurons, are clearly visible on the map [see Figs.~\ref{fig:SOM}(a) and \ref{fig:SOM}(d)], which reflects the topology of the target space. The presence of dead neurons is due to the fact that falling close to the middle of the region between clusters, the neuron weight vector, in full agreement with Eq.~(\ref{eq:SOM_training}), undergoes a multidirectional displacement during the learning process, caused by alternate attraction from the neighbors that have fallen into different clusters. Such oscillations compensate each other on average, as long as the distance between the clusters is significantly larger than their amplitude, but as the temperature increases, the clusters become closer and this condition breaks down making the position of the weight vector unstable [see Figs.~\ref{fig:SOM}(b) and \ref{fig:SOM}(e)]. The probability of attraction to one of the clusters increases, which leads to a sharp decrease in the number of dead neurons and allows us to consider it as a characteristic parameter that specifies the phase of the system [see Figs.~\ref{fig:SOM}(c) and \ref{fig:SOM}(f)], i.e., the critical temperature (for a more rigorous mathematical analysis, see Supplemental Material, Sec.~D \cite{supp}). From Figs.~\ref{fig:SOM}(c) and \ref{fig:SOM}(f) we conclude that the Curie temperature $T_c\approx220$~K for the 2DFM and $T_c\approx915$~K for bcc Fe, which is in excellent agreement with the results obtained from the Binder cumulant analysis \cite{supp} and also reproduces experimental values of bcc Fe. The sharp change of the SOM neural activity at the critical temperature makes applications of this method more precise as compared to PCA \cite{supp} and easier than cumulant method theory, because the simulation of the only one system of representative size is demanded; as one can see from  Figs. 2(c) and 2(f), plots of neural activity are almost the same for biggest lattices, representing the fact that obtained results are independent of the model size, since it becomes big enough (a brief explanation is given in the Supplemental Material, Sec.~B \cite{supp}). Whereas in the Binder cumulant technique, a set of different sized systems are needed. 

{\it Conclusions.} In this Rapid Communication we proposed an approach for phase detection in systems with second-order transition, where the state is described by a large number of vectors. The method is based on constructing a special multidimensional target space with phase-related topology and an unsupervised learning algorithm of SOMs that is used to determine and visually observe a phase transition. We applied the method to characterize the phase transition and for calculating the critical temperature of a two-dimensional ferromagnet on a square lattice and bcc Fe. Our findings reveal an excellent agreement, being compared with results obtained with the conventional technique of the Binder cumulant theory. As opposed to the cumulant method theory that requires one to scale up the size of a system, the here suggested method allows one to make realistic predictions having Monte Carlo simulations for one copy of the system of a certain size only. The latter makes it possible to further utilize the method for various  applications in statistical physics and condensed-matter systems, not only in magnetism but also, for example, for order-disorder transitions in alloy theory. A possible extension of the method proposed here is to provide a deeper understanding of short-range order around phase transitions, where experimental data exists, e.g., based on muon spin spectroscopy. We believe the proposed method can be generalized for the problems of purely quantum-mechanical nature as long as the elements of the corresponding system are described by certain vectors in multidimensional space, while the phase transition is associated with a change in their angular distribution.

{\it Acknowledgments.} A.A.S. and D.Y. acknowledge support from the Russian Science Foundation Project No. 17-12-01359. V.K.K. acknowledges support from Horizon2020 RISE project CoExAN. O.E. acknowledges support from the Swedish Research Council (VR), the Knut and Alice Wallenberg Foundation (KAW), the Foundation for Strategic Research (SSF), the Swedish Energy Agency, eSSENCE, and STandUPP.

\bibliographystyle{apsrev4-1}

\newpage

\onecolumngrid
\setcounter{page}{0}
\setcounter{table}{0}
\setcounter{section}{0}
\setcounter{figure}{0}
\setcounter{equation}{0}
\renewcommand{\thepage}{\Roman{page}}
\renewcommand{\thesection}{S\arabic{section}}
\renewcommand{\thetable}{S\arabic{table}}
\renewcommand{\thefigure}{S\arabic{figure}}
\renewcommand{\theequation}{S\arabic{equation}}
\cleardoublepage
\vfill\eject
\thispagestyle{empty}
    
\section*{Supplemental Material}
    
\subsection{Monte Carlo simulations}
We consider a two-dimensional exchange model on a square lattice (2DFM) with the Hamiltonian,
\begin{equation}
    H_\mathrm{2DFM}=-\frac{J}{2}\sum\limits_{\langle\bm{r},\bm{r}'\rangle}\bm{S}_{\bm{r}}\cdot\bm{S}_{\bm{r}'}-K_\mathrm{anis}\sum\limits_{\bm{r}}(S_{\bm{r}}^z)^2,
\end{equation}
where the summation between the nearest neighbors $\langle\bm{r},\bm{r}'\rangle$ is implied, the exchnage coupling strength $J=1$ mRyd, and the parameter of easy-axis anisotropy $K_\mathrm{anis}=0.2$ mRyd.
\begin{figure}[h!]
    \includegraphics[width=\linewidth]{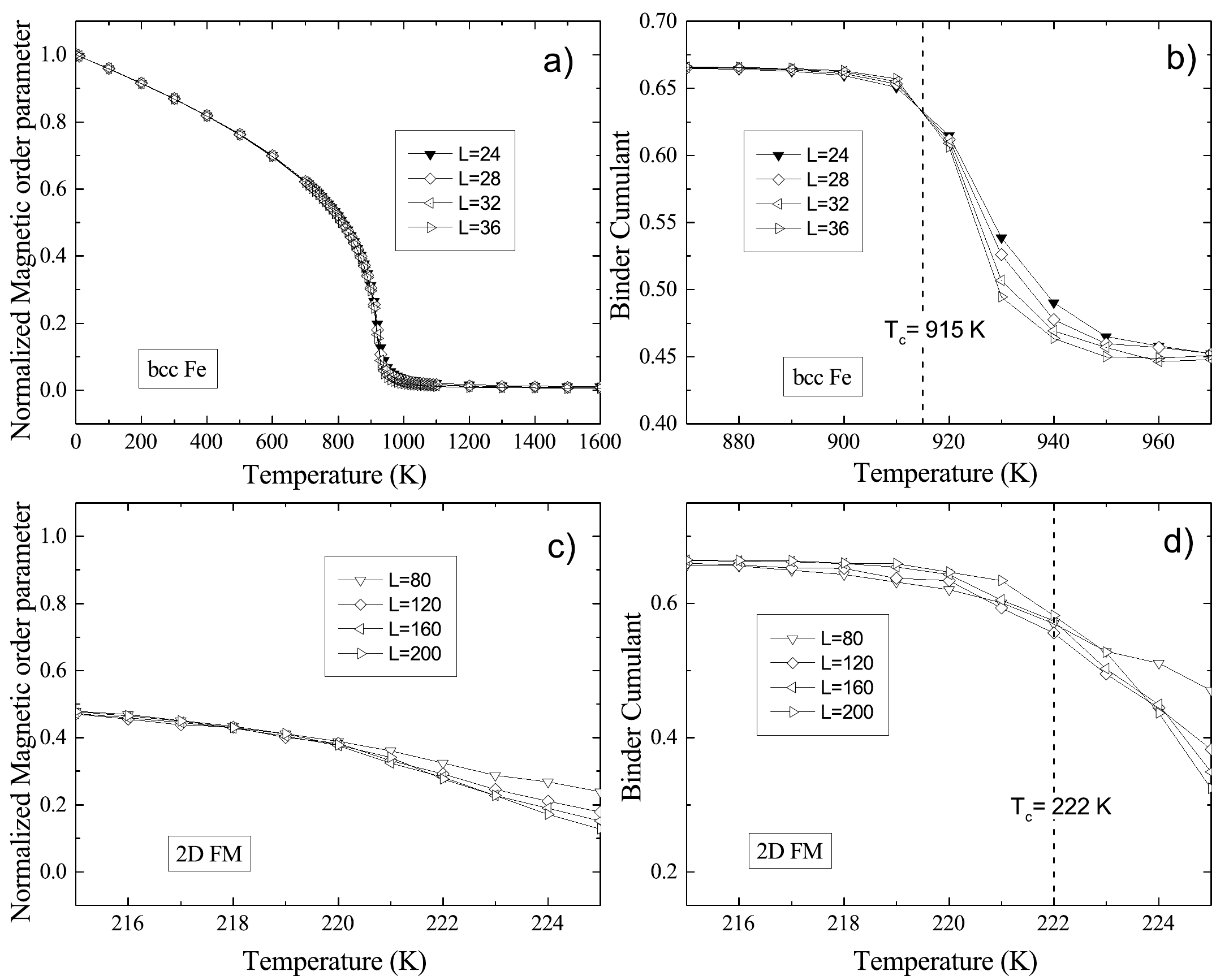}
    \caption{
    Magnetic phase diagrams obtained from Monte Carlo simulations, indicating that both the 2DFM and bcc Fe have second order phase transitions. The magnetic order parameter $M(T)$ for bcc Fe (a) and 2DFM (c). The Binder cumulant $U(T)$ for bcc Fe (b) and 2DFM (d). The dotted line passes through the fixed point of different $U(T)$ corresponding to the different lattice sizes (different $L$), indicate the phase transition temperature.}
    \label{fig:magn_and_binder}
\end{figure}
The Hamiltonian that corresponds to three-dimensional magnetism in the body-centered cubic structure can be modelled by,
\begin{equation}
    H_\mathrm{bcc}=\sum\limits_{\bm{r}}\sum\limits_p J_p\bm{S}_{\bm{r}}\cdot\bm{S}_{\bm{r}+\bm{p}},
\end{equation}
where the exchange coupling parameters $J_1=1.3377$, $J_2=0.7570$, $J_3=-0.0598$, and $J_4=-0.0882$ mRy are obtained from first principles electronic structure calculations, and the vector $\bm{p}=n_1\bm{e}_1+n_2\bm{e}_2+n_3\bm{e}_3$, on condition that $\bm{e}_1$, $\bm{e}_2$, and $\bm{e}_3$ stand for the Bravais lattice vectors of the bcc lattice, while the integers $n_1+n_2+n_3=p$. It is noteworthy that for both 2DFM and bcc Fe $\bm{S}_{\bm{r}}$ represents a vector of unit length.

The results of Monte Carlo simulations are shown in Fig.~\ref{fig:magn_and_binder}, with the graphs of the magnetic order parameter $M(T)$ displayed in Fig.~\ref{fig:magn_and_binder}(a) and (c) and of the Binder cumulant $U(T)$ shown in Fig.~\ref{fig:magn_and_binder}(b) and (d). We used simulation cells with edge lengths $L=80$, 120, 160, and 200, corresponding to a total number of spins $N=6400$, 14400, 25600, and 40000  respectively for 2DFM. While for bcc Fe simulations cells of edge lengths $L=24$, 28, 32, and 36, corresponding to a total number of spins $N=27648$, 43904, 65536, and 93312 respectively have been utilized. The phase transition between ferromagnetic and paramagnetic phases of matter for both 2DFM and bcc Fe is clearly visible. The critical temperature for the 2DFM is estimated to be around $T_c\approx$ 222~K and for bcc Fe $T_c\approx 915$~K.

\subsection{Clustering detection by PCA}
\begin{figure*}[!t]
    \includegraphics[width=\linewidth]{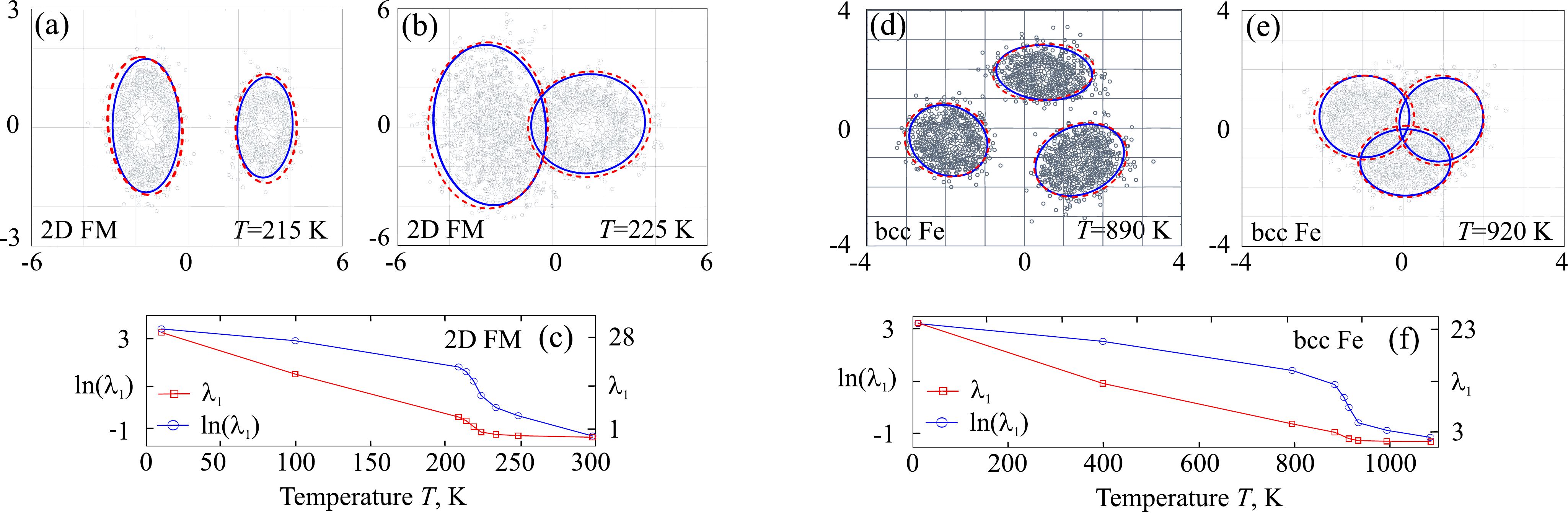}
    \caption{Visualization of clustering in the target space by PCA and evolution of data dispersion. The groups of vectors -- grey dots at (a), (b) for 2D FM, $L=120$; (d), (e) for bcc Fe, $L=28$ -- representing projections of sites spins, move towards each other with the increase of temperature from below ($T=215$~K for 2DFM, and $T=890$~K for bcc Fe) to above ($T=225$~K for 2DFM, and $T=920$~K for bcc Fe) the critical temperature. At the same time, the first (maximum) eigenvalue of the covariance matrix $\lambda_1$ for 2DFM (c), and bcc Fe (f), visualizing the dispersion of data in the target space, decreases, but its evolution has no significant change across the phase transition (the logarithmic scale is added to highlight the behavior of the $\lambda_1$ in the vicinity of the critical temperature). Solid blue and dotted red lines, surrounding clusters, show the elliptic fit, containing 95\% of vectors, for different lattice sizes: at (a), (b) blue is for $L=120$, red is for $L=200$; at (d), (e) blue is for $L=28$, red is for $L=36$.}
    \label{fig:PCA}
\end{figure*}

The principle component analysis (PCA), one of the main methods of data dimensionality reduction, has been designed to minimize the loss of information and found wide applications in many areas of research. Intuitively, the PCA can be thought of as fitting an $n-$dimensional ellipsoid to the data, and projecting this data onto the subspaces formed by ellipsoid axes -- this is equivalent to finding the set of orthogonal directions along which the variance of data is maximal. Having a set of observations $\bm{X}_i=(x_{1i},...,x_{mi})^T$ of a system with $m$ degrees of freedom, we can characterize it by its empirical covariance matrix $C=[c_{ij}]$,
\begin{equation}
    \label{eq:Cov_matrix}
    c_{ij}=\frac{1}{m-1}\sum_{l=1}^{m} (x_{li}-\overline{\bm{X}}_i)(x_{lj}-\overline{\bm{X}}_j).
\end{equation}
In particular, PCA aims at determining the eigenvectors of this matrix to perform a lower dimensional projection. Having constructed the target space (see the main text for details) we are to visualize the evolution of clustering by projecting out the target space onto the subspace of the first two principal components. In the ferromagnetic phase we explicitly observe well-separated clusters [Figs.~\ref{fig:PCA}(a) and (d) for 2DFM and bcc Fe respectively], having the tendency of getting closer when raising up the temperature and eventually merging together for temperatures higher than the critical one [Figs.~\ref{fig:PCA}(b) and (e) for 2DFM and bcc Fe respectively]. Meanwhile, a common technique consisting in comparing the first eigenvalue [Figs.~\ref{fig:PCA}(c) and (f)] and the magnetic order parameter in this case is not applicable in view of the special geometry of the target space in which the dispersion of the data distribution does not have such an explicit physical meaning.

Meanwhile, the PCA can provide an illustrative explanation for the fact that the results obtained within SOM methodology are independent of the size of a spin lattice (as shown in Fig.~2 in the main text). It can be attributed to the fact that as the size of the lattice increases, the boundaries of the clusters in the target space become sharper, while their sizes grow insignificantly. However, when a particular concentration of vectors in the tiny region close to the border is reached, the accuracy is purely determined by the step of Monte Carlo simulations, and cannot be considerably improved with increasing lattice size, which only more sharply outlines their boundaries -- as one can see from [Figs.~\ref{fig:PCA}(a),(b),(d),(e)] elliptical fits, containing 95\% of clusters vectors, are almost identical for given lattice sizes before and above the critical temperature.

\subsection{Choosing the map size}
\begin{figure*}[!t]
    \includegraphics[scale=0.75]{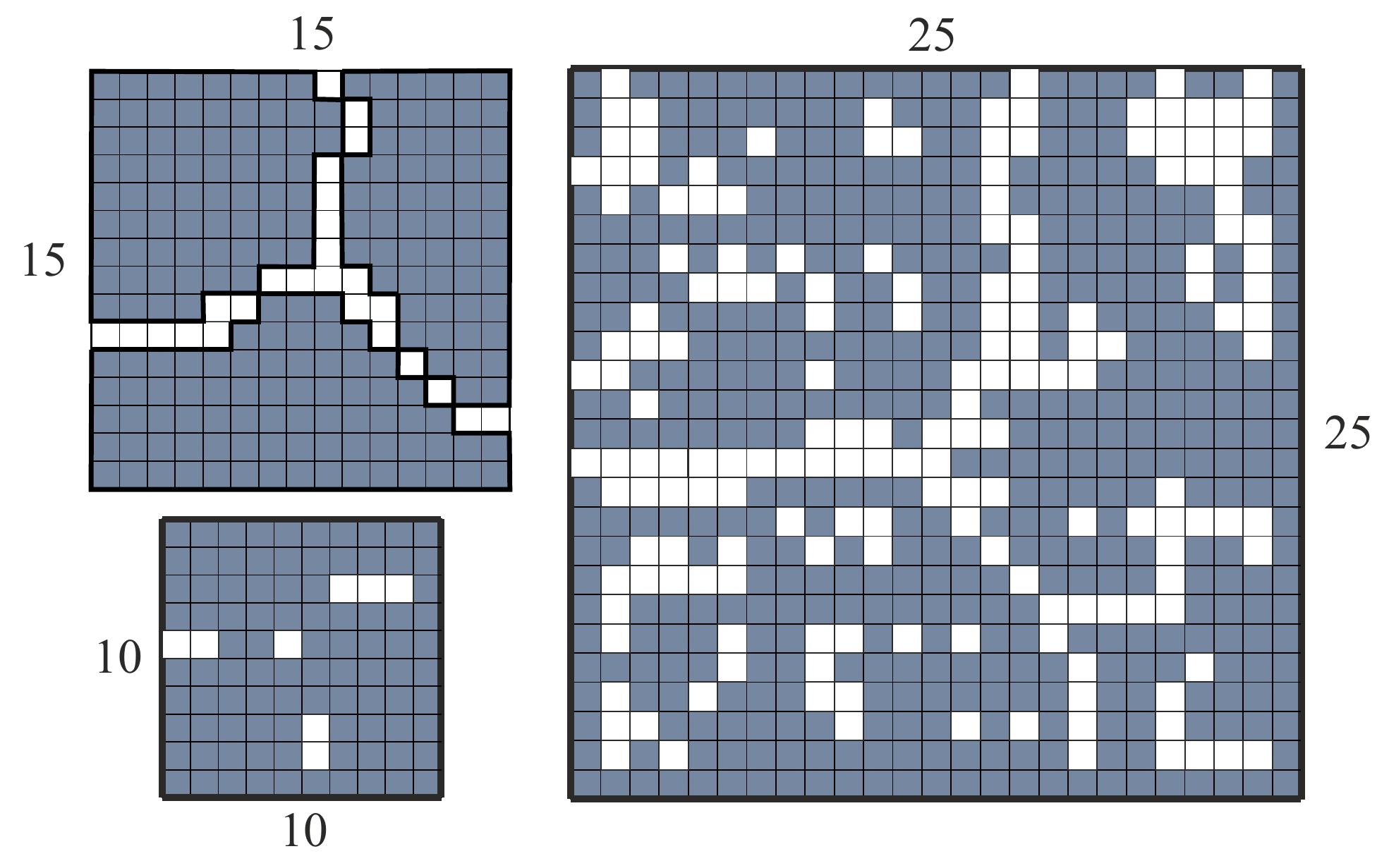}
    \caption{Maps of different size, representing projection of the bcc Fe target space at $T=890$ K. Too many input vectors per neuron are present resulting in no cluster formation ($10\times10$ map); unstable cluster structure due to insufficient amount of input data ($25\times25$ map); the optimal configuration with sharp and visible  boundaries is achieved ($15\times15$).}
    \label{fig:MAP_sizes}
\end{figure*}

The way how to determine the size of a map can be justified as follows. If the size of a map is too small there are too many input vectors per neuron, resulting in either no dead neurons at all, or very few of them, so that regardless of the phase no structure can be visualized [Figs.~\ref{fig:MAP_sizes}, e.g., $10\times10$]. Otherwise, if the size of the map is too large the amount of input data per neuron is too small, leading to a large amount of dead neurons weakly correlating to the phase of a system [Figs.~\ref{fig:MAP_sizes}, e.g., $25\times25$]. The optimal size of a map that correctly displays the cluster structure of the target space and is sensitive to phase detection is typically chosen in between these two limits [Figs.~\ref{fig:MAP_sizes}, e.g., $15\times15$ that has been used in our work], and can be found by increasing the size of a certain small map until the stable cluster picture is visible (usually, there is a range of map sizes, giving the same picture and results). It might be of use to firstly perform PCA to get a proper feeling of the target space geometry.

\subsection{Mathematical grounds of SOM phase identification}
    
In order to analyze the mechanisms making phase detection with SOMs possible, we proceed with an analysis of the training process. Let us fix the map node $\bm{r}$, two of its neighbours $\bm{r}_1$ and $\bm{r}_2$, and assume that the weight vector $\bm{w}(\bm{r})$ is located between two clusters $C_1$ and $C_2$. As a result of the training process, such a node may become a dead neuron, and it is natural to understand how the probability of such an event depends on a system phase. If both neighbouring weight vectors $\bm{w}(\bm{r}_1)$ and $\bm{w}(\bm{r}_2)$ are in the same cluster, they make $\bm{w}(\bm{r})$ to move towards this cluster, increasing the chance of activation straight forwardly. The more interesting case is when the data vectors $\bm{d}_1$ and $\bm{d}_2$, to which the weights of $\bm{r}_1$ and $\bm{r}_2$ nodes are closest, are located in different groups. To analyze this situation we rewrite and reduce Eq.~(\ref{eq:SOM_training}) of the main text with the fixed $\bm{r}$ to,
\begin{equation}
    \label{eq:SOM_training_2}
    \bm{w}_{n+1}=\bm{w}_{n}[1-\alpha_n\theta_n(\bm{r}')]+\alpha_n\,\theta_n(\bm{r}')\,\bm{d}_m,
\end{equation}
and assume that $\bm{w}_0=(\bm{d}_1+\bm{d}_2)/2$. The latter means that initial value of $\bm{w}(\bm{r})$ is located right in the middle between $\bm{d}_1$ and $\bm{d}_2$, and represents the restriction of our intuitive assertion, that the weights of dead neurons should be somewhere in the middle between clusters. If the system is in the ferromagnetic phase and Eq.~(\ref{eq:phase_cond}) of the main text is satisfied, the probability for the weight of the node $\bm{r}$, located between clusters, to be the closest to a certain data vector $\bm{d}_0$ can be roughly estimated as $\propto$ $A^{-N}$, where $A$ is a constant. Thus, in the majority of cases, the evolution of  $\bm{w}(\bm{r})$ is only determined by its neighbours $\bm{r}_1$ and $\bm{r}_2$, which makes it possible to write down equations of the training process in the form,
\begin{equation}
    \label{eq:SOM_training_ferro}
    \bm{w}_{n}=\bm{w}_{n-1}(1-\alpha_{n-1}\theta_{n-1})+\alpha_{n-1}\theta_{n-1}(2\bm{w}_0-\bm{d}_1), \quad
    \bm{w}_{n+1}=\bm{w}_{n}(1-\alpha_n\theta_n)+\alpha_n \theta_n \bm{d}_1,
\end{equation}
where we omitted the argument in $\theta_{n}(\bm{r}')$, as this function depends on the relative distance $|\bm{r}'-\bm{r}|$, which are equal for $\bm{r}_1$ and $\bm{r}_2$. Doing the sum in Eqs.~(\ref{eq:SOM_training_ferro}), we derive
\begin{equation}
    \label{eq:SOM_training_ferro_sum}
    \bm{w}_{n+1}=-\alpha_n\theta_n\bm{w}_{n}+\bm{w}_{n-1}-\alpha_{n-1}\theta_{n-1}\bm{w}_{n-1} +2\bm{w}_0\alpha_{n-1}\theta_{n-1}+\bm{d}_1(\alpha_n\theta_n-\alpha_{n-1}\theta_{n-1}),
\end{equation}
which in the limiting case $n\to\infty$ results in,
\begin{equation}
    \label{eq:SOM_training_ferro_limit}
    \lim_{n \to \infty} \bm{w}_{n}=\bm{w}_0,
    \end{equation}
meaning that in the ferromagnetic phase weights of dead neurons retain their values. However, by increasing the temperature, provided that the distance between clusters becomes comparable with the amplitude of the node weight oscillations,
\begin{equation}
    \label{eq:SOM_training_para_cond}
    \text{dist}(C_1, C_2) \sim \alpha \max[\text{diam}(C_1), \text{diam}(C_2)],
\end{equation}
leads to that the dead neuron weight acquires a non-zero probability to approach the cluster and it becomes closest to one of the data vectors. If this is the case, the equations of the training process [Eqs.~(\ref{eq:SOM_training_ferro})] should be supplemented with
\begin{equation}
    \label{eq:SOM_training_para_add}
    \bm{w}_{n-1}=\bm{w}_{n-2}(1-\alpha_{n-2})+\alpha_{n-2} \bm{d}_0,
\end{equation}
where $\theta_n(\bm{r})=1$, and $\bm{d}_0$ is the vector to which $\bm{w}(\bm{r})$ is closest. Solving Eqs.~(\ref{eq:SOM_training_ferro}) together with (\ref{eq:SOM_training_para_add}) gives rise to, 
\begin{equation}
    \label{eq:SOM_training_para_prelimit}
    \lim_{n\to\infty}\bm{w}_{n}=\frac{2\bm{w}_0\lim_{n\to \infty}\theta_n+\bm{d}_0}{2\lim_{n\to\infty}\theta_n+1} = \bm{d}_0,
\end{equation}
since $\lim_{n\to\infty}\theta_n = 0$ owing to the training algorithm. Thus, as soon as the weight of a dead neuron becomes close enough to a certain vector in the target space, its position becomes unstable and the node turns into activated one, which means that the amount of dead neurons dramatically decreases when clusters are close to each other, which is the situation in the vicinity of the critical temperature.
    
\end {document}